# Ontological metaphors for negative energy in an interdisciplinary context

Benjamin W. Dreyfus, Benjamin D. Geller, Julia Gouvea,
Vashti Sawtelle, Chandra Turpen, and Edward F. Redish

*Department of Physics, University of Maryland, College Park, MD 20742*

***Abstract.*** Teaching about energy in interdisciplinary settings that emphasize coherence among physics, chemistry, and biology leads to a more central role for chemical bond energy. We argue that an interdisciplinary approach to chemical energy leads to modeling chemical bonds in terms of negative energy. While recent work on ontological metaphors for energy has emphasized the affordances of the substance ontology, this ontology is problematic in the context of negative energy. Instead, we apply a dynamic ontologies perspective to argue that blending the substance and location ontologies for energy can be effective in reasoning about negative energy in the context of reasoning about chemical bonds. We present data from an introductory physics for the life sciences (IPLS) course in which both experts and students successfully use this blended ontology. Blending these ontologies is most successful when the substance and location ontologies are combined such that each is strategically utilized in reasoning about particular aspects of energetic processes.

## I. INTRODUCTION

Energy is a central concept in physics, chemistry, and biology, and has been widely promoted [1] as a way to connect physics and chemistry to biology. Yet the concept of energy can be fractured for students along disciplinary lines.[2,3] Chemical energy (energy changes associated with chemical bonds and reactions) is essential in biology and chemistry [4], and rarely has a central role in introductory physics courses. However, introductory physics courses that seek deeper interdisciplinary coherence with chemistry and biology are now integrating chemical energy into their treatment of energy.[5] We argue below that one element of building this interdisciplinary coherence around chemical energy is reasoning about negative energy. However, we note that this would be less essential in other introductory physics curricula.

Negative energy has been documented as an area of difficulty for students.[6,7] In this paper, we draw on an ontological metaphor perspective to suggest why this concept is difficult, and use a dynamic ontologies model to illustrate ways that experts and students can reason productively about negative energy.

In Part II, we review the PER literature on ontological metaphors, particularly as applied to energy. We focus on two metaphors for energy: substance and location. In Part III, we discuss the concept of negative energy: why it is pedagogically necessary for our interdisciplinary context, and how it has been a source of confusion. In Part IV, we argue that the exclusive use of the substance metaphor for energy is untenable for an interdisciplinary context that relies on negative energy, and present examples of the productive use of a blended substance/location ontology. In Part V, we present a case study of one group problem-solving task on energy at molecular scales, and analyze student reasoning about negative energy with a focus on ontological metaphors. In Part VI, we discuss the implications for research and for instruction, including suggesting the instructional value of coordinating multiple ontologies, and proposing future directions for research beyond this paper's narrow context.

## II. THEORETICAL FRAMEWORK

### A. Ontologies and conceptual metaphors in physics education

Our analysis is based in the conceptual metaphor theory developed by Lakoff and Johnson [8]. This theory elucidates the metaphors we use, based in our physical experiences in the world, when we think and talk about abstract ideas. These include ontological metaphors, which Lakoff and Johnson define as "ways of viewing events, activities, emotions, ideas, etc., as entities and substances." For example, "He cracked under pressure" is an instance of the *The Mind Is A Brittle Object* metaphor.

Another strand of research on ontologies in learning physics is based in the work of Chi and colleagues [9–12]. They build on the theory of Keil [13], which posits that all entities in the world can be placed into a hierarchy of ontological categories, and apply this theory to science concepts, using Matter, Processes, and Mental States as the primary ontological categories.

According to Chi et al.'s theory, each physical entity has a correct ontology, and physics misconceptions are the result of attributing an incorrect ontology to a concept. While we do not share this theoretical perspective, we draw on Chi et al.'s methodology of identifying ontologies that students (and experts) use by analyzing the predicates that they use and associating these predicates with ontologies.

Brookes and Etkina [14] synthesize the conceptual metaphor framework and the ontological categories framework. They follow Chi et al. in placing each physics concept into an ontological category based on expert understanding of physics (a lexical ontology), but they also identify instances when students and experts invoke other ontologies for a given concept. When these ontologies do not match the lexical ontology, they identify this as a metaphor.

Gupta et al. [15] respond to Chi et al.'s "static ontologies" model, and show that both novices and experts can place the same physics entity in multiple ontological categories, and that this ontological categorization is context-dependent. They show furthermore that using multiple complementary ontologies for the same concept in different contexts can be productive. We extend this dynamic ontologies model to cases in which multiple ontological categories are used for the same entity within the same episode.

## B. Ontological metaphors for energy

In recent years, a popular theme in the physics education research literature has been the use of ontological metaphors for energy: conceptual metaphors that express "what kind of thing energy is."[16]

Scherr et al. [16] identify three ontologies for energy found in student and expert discourse:
• *Substance:* energy as "stuff" contained **in** objects
• *Stimulus:* energy **acts on** objects
• *Vertical location:* objects are **at** higher or lower energies, by analogy to gravitational energy.

They note that "the stimulus metaphor is not common in expert physicists' discourse about energy," and likewise here we focus primarily on the substance and vertical location metaphors, both of which are commonly used by expert physicists.

All three of these ontologies are metaphorical according to Brookes and Etkina's definition [14]: Energy is an abstract concept that is not "actually" a substance or a location according to canonical physics understanding. Therefore, in this particular domain, we are justified in referring to "ontologies" and "metaphors" largely interchangeably in this paper (in keeping with other literature in this area), even if they are not always equivalent in other cases.

We should clarify here the distinction between the substance and location ontologies for energy. Amin's [17] conceptual metaphor analysis of energy identifies attributes of energy with elements of Lakoff and Johnson's [18] Object Event-Structure and Location Event-Structure metaphors. Both of these fundamental metaphors create spatial mappings for events, but the Location Event-Structure metaphor identifies events with locations (e.g. "He went into a depression"), and the Object Event-Structure metaphor identifies events with objects (e.g. "I have a headache"). It may appear that these metaphors correspond to the substance and location ontologies respectively, but this correspondence is not accurate, because our focus is on what the metaphors imply about what energy is, rather than about the metaphors themselves. The Object Event-Structure metaphor does indeed correspond to the energy-as-substance ontology; this includes possession language about "having" energy. However, different uses of the Location Event-Structure metaphor may correspond to either the substance or the location ontology for energy. As one example of the Location Event-Structure metaphor, Amin includes energy being "in" some form. We would still classify this as the substance ontology, because the energy is "in" the metaphorical "location" (and being at a location is a predicate associated with a substance) rather than the energy itself being the location. In another context, Amin writes "Here again we find the Location Event Structure conceptual metaphor, but now with a figure/ground reversal. Energy transformation was construed in terms of this metaphor. In that case, energy was construed as an object moving from one location to another. Here, in contrast, we find that energy state is the location and objects move with respect to it." This is the context that we identify as the energy-as-location ontology.

When we discuss the location ontology, we are also not referring to situations where the energy of an object depends on the object's spatial location. In those situations, the location of the object is not a metaphor, but a physical property. While the energy may depend on the location, the energy is independently described by some ontology, which may or may not also be the location ontology. (This can be a source of confusion for both students and researchers in understanding potential-energy-vs.-position graphs, because the horizontal axis on those graphs represents spatial location, while the vertical axis, representing energy, can be interpreted as a metaphorical location. As we discuss below, this can also help activate productive conceptual resources.)

After describing three common ontologies for energy, Scherr et al. [16] go on to focus on the substance ontology, making the case for its pedagogical ad-vantages and detailing how it can be used in instruction. Brewe [19] takes a similar approach, also focusing on the energy-as-substance metaphor as a central framework for an introductory physics curriculum. Lancor [20] examines the use of conceptual metaphors for energy in all three disciplines, and also focuses on the substance metaphor in its various manifestations.

All of these recent papers share a theoretical commitment to dynamic ontologies.[15] This stands in contrast to the "static ontologies" view [9] that there is one correct ontological category corresponding to each entity, and misconceptions arise from ontological miscategorizations. Thus, when Scherr et al. and Brewe advocate for emphasizing the substance ontology in instruction, they are not claiming that the substance ontology is the "correct" ontology for energy; rather, their claims are based on the pedagogical affordances of this metaphor. These affordances include supporting the ideas that energy is conserved, can be located in objects, is transferred among objects [16], and is unitary (i.e., there is only one type of energy) [19] and/or can change form.[20]

However, they concede that one place where the substance metaphor encounters difficulties is the representation of negative energy, since a substance cannot ordinarily be negative. Scherr et al. resolve this concern with "the realization that potential energy depends not only on the system of mutually interacting objects but also on a reference point." In other words, it is possible to choose a reference point such that the potential energy of the system of interest is always positive, enabling the use of the substance metaphor. In Brewe's Modeling Instruction course, energy is first visually represented with pie charts, which emphasize conservation and unitarity. This representation breaks down when attempting to incorporate negative energy, and this provides the motivation to replace pie charts with bar charts [21], which can represent negative energy. However, it is less clear that bar charts embody the substance metaphor in the way that pie charts do, or how negative bars fit into the structure of this metaphor. The case study in section V will present examples of students reasoning about positive and negative energies with the bar chart representation, and illustrate that they are not necessarily stably associated with a single metaphor. In sections III and IV we will discuss the negative energy issue and suggest a solution consistent with student and expert data and with the dynamic ontologies perspective.

## III. INTERDISCIPLINARITY AND NEGATIVE ENERGY

Our research in this area is in the context of developing the NEXUS/Physics course [22,23], an introductory physics course[1] for undergraduate biology students that is focused on building interdisciplinary coherence between physics, biology, and chemistry. In a traditional introductory physics course, the energy curricular unit focuses on mechanical energy: kinetic energy and macroscopically detectable potential energies (usually gravitational and elastic). "Chemical energy" is most typically treated as a black box (to account for where the missing mechanical energy went) if at all.[4] This approach comes up short for biology students, because most energy relevant in biological systems is chemical energy (i.e. energy changes associated with chemical bonds and chemical reactions), and so the traditional physics sequence does not give them the appropriate tools to analyze energy in biological situations.

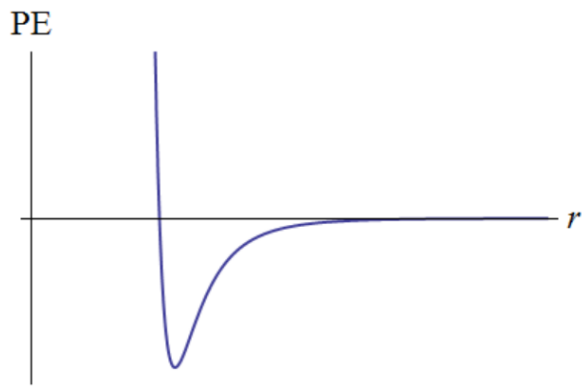

**FIGURE 1.** The Lennard-Jones potential, approximating the interaction between two atoms.

Therefore, chemical energy is a core component of the NEXUS/Physics course's treatment of energy[5], following other physics courses for the life sciences.[24] Electric forces and electric potential energy are moved up to the first semester and used to model (qualitatively) the potential for a system of two interacting atoms (Figure 1). This leads to a description of chemical bonds in terms of electric potential energy and other constructs that connect to the overall conceptual framework of physics.

The concept of negative energy is essential to this model of chemical bonds. When two atoms are bound, their energy is less than the energy of the same atoms if they were unbound. If the energy of unbound atoms is taken to be zero, then the energy of the bound atoms is negative. Unlike models of gravitational potential energy (*mgh*) that are common in introductory physics courses, the "zero" point of potential energy in this model is not arbitrary. Zero potential energy has a specific physical meaning here: the energy when the atoms are far enough apart that they are not interacting. Shifting the zero point below the strongest bond in the system to make all energies positive (in order to preserve the substance ontology) would mean that adding new molecules to the system (which have the capacity to form additional bonds) would require shifting the zero again, with no limit. Modeling bound atoms with negative energy contributes substantial conceptual clarity by allowing for a common "zero" point in the absence of interaction. Therefore, when chemical energy is a central piece of the overall energy picture, the representational tools in use need to be set up so that negative energy is accessible from the beginning.

---

[1] See http://nexusphysics.umd.edu .

While there are sound conceptual reasons for the use of negative energy to model chemical bonds in this context, we know that negative energy has also been shown to be a subject of confusion for students. Stephanik and Shaffer [6] document the belief that potential energy cannot be negative, as well as the belief that kinetic energy cannot exceed total energy. (This latter belief may also have roots in the substance ontology; if the total energy is a pie, it is inconceivable that one slice of the pie could be larger than the entire pie.) Lindsey [7] shows a tendency to look only at the magnitude of the potential energy, and therefore to conclude that a system of two (electrostatically or gravitationally) attracting objects has greater potential energy when the objects are closer together. While these concerns may weigh against the instructional use of negative energy, they may be mitigated by ontological choices in reasoning about energy. Specifically, as we will discuss in the next section, reasoning about negative energy with the location ontology may bypass these difficulties.

## IV. BLENDING THE ONTOLOGIES

While other authors operating in different instructional contexts have argued for the primary use of the energy-as-substance ontology, our student population and curricular goals lead us to a different cost-benefit analysis. Scherr et al. [16] are exploring these questions in the context of a professional development program for K-12 teachers, and Brewe's [19] Modeling Instruction course is for undergraduates from all the science and engineering majors. Neither context demands the same special concerns that are occasioned by our interdisciplinary context that attempts to form deep connections between physics and biology. The centrality of negative energy in modeling bonding and chemical reactions means that an exclusive substance ontology for energy is untenable. (Paradoxically, it is not only straight "physics" contexts that are able to sufficiently black-box chemical energy to treat it as a positive substance. Straight biology contexts frequently do the same. It is the interaction between physics and biology, and the use of physics constructs to describe phenomena relevant to biology, that necessitates opening up this black box and engaging with negative energy.)

The energy-as-vertical-location metaphor is better suited for energies that can be positive or negative: Extending the substance ontology to negative quantities requires complicated maneuvering (e.g. defining a negative substance that cancels out when it combines with the positive substance). However, it is no more conceptually difficult to be at a location "below" zero than at a location "above" zero. The location ontology for energy is also in common usage among expert physicists, such as in the potential well metaphor [14].

However, it is hard to imagine a comprehensive picture of energy that is based exclusively on the location ontology. The location metaphor succeeds at capturing some important aspects of energy: energy is a state function (i.e., the energy of a system is independent of the path that the system took to reach that state); energy can be positive or negative; changes in potential energy are more physically meaningful than the actual value of potential energy (not obvious in the substance metaphor, in which the value of potential energy appears to have physical meaning); intuitions based on gravitational potential energy about the relationship between energy and force (and embodied experience about up and down) can be applied to other non-gravitational energies. But there are other aspects that the location metaphor represents less effectively: interactions and energy transfer among objects in a system; energy is conserved.

The use of these two metaphors for negative numbers is explored extensively in the mathematics education literature, though not in the same language we use here. Ball [25] writes about teaching negative numbers to elementary students, and uses two primary models: a building with floors above and below ground (analogous to the vertical location metaphor), and money and debt (analogous to the substance metaphor). The students in Ball's study had greater difficulty with the money and debt representation. Streefland [26] and Linchevski and Williams [27] contrast substance metaphors for negative numbers (positive and negative cubes) with thinking about positive and negative numbers as processes or changes in some other quantity (people getting on and off a bus).

Though neither the substance nor the location ontology for energy is adequate on its own for the reasons outlined above, combining the two addresses these shortcomings. We suggest that the framework of conceptual blending [28] is appropriate to describe this combination of ontologies. The authors are currently developing a more rigorous analysis of why this constitutes blending the ontologies (rather than switching between two distinct ontologies) which will appear in a future publication.

The blended substance/location ontology for energy is common among expert physicists. This is illustrated by the following classroom transcript from a physics professor teaching the NEXUS/Physics course. We analyze the transcript data by coding for predicates [11] associated with the substance and location ontologies. The use of the energy-as-substance metaphor is underlined, and the use of the energy-as-location metaphor is in **bold**. This coding excludes language (such as "get them back apart") that refers to the *spatial* location of the atoms, since that location is literal and is not a metaphor for energy.

> *If the two atoms are apart and form a bond, they **drop down to here** and release that much energy. And because that's **where they are, at that negative energy**, that's equal to the energy you have to put in to get them back apart. So it's just about **where you're going**, that when you're forming a bond, you're **dropping down**, and if you come in **at this energy** you gotta get rid of this much. But if **you're down here** and you want to **get back up to here**, you gotta put in this much.*

Here, the substance and location ontologies are combined in a productive way, and the professor fluidly moves between these metaphors within a single sentence. The blended ontology is consistent: the energy of the system of atoms is described as a vertical location, and changes in the energy of the system are described as a substance (that enters or leaves the system). There is nothing extraordinary about this quotation; it illustrates a standard way that expert physicists talk about energy, especially in atomic and molecular contexts. Another typical example is found in *The Feynman Lectures on Physics* [29]: "If an atom is initially **in one of these 'excited states,'** … sooner or later **it drops to a lower state** and radiates energy in the form of light."

This blending can also be productive for students. A well-documented issue in biology and chemistry education is the student difficulties around "energy stored in bonds."[3] The causes of this problem can be traced to multiple sources, but the inappropriate application of a substance ontology for energy may be partially responsible. The substance ontology supports a metaphor in which a bond is a piñata containing "stuff," and the stuff (energy) is released when the bond is broken. One student, Anita[2], explained in class that she used to think about bonds this way: "whenever chemistry taught us like exothermic, endothermic, … I always imagined like the breaking of the bonds has like these little molecules that float out." She was using this metaphor "until I drew … the potential energy diagram, and that's when I realized, to break it you have to put in energy." In a follow-up interview in which she was reflecting on this specific discussion in class, Anita explained her use of the potential energy graph (with the substance and location predicates once again coded in the transcript):

> *What I imagine it is, to get it to break, you need to put in energy. So to **get up the hill**, you need to input energy ... Say that you're **bicycling up the hill**. You need energy to put it in, that's what breaks the bond, but to bring them back together, it's released. So I just think of—when you're **falling down**, if you're **going down a hill with a bike**, you're not putting in energy to the pedals, but yet your pedals keep going so there's energy released.*

According to Anita's self-report, her initial exclusive use of the substance metaphor led her to claim incorrectly that energy is released when bonds are broken. In this interview clip, we see Anita using the location metaphor to leverage intuitions about gravity in a non-gravitational context. Switching to a blended substance/location ontology has helped her develop a more correct understanding of chemical bond energy.

The data in this section are "clean" examples of blending the substance and location ontologies for energy, representing a way of thinking about energy that a student or expert has already found productive. In the next section we examine some "messier" examples, in which this blend arises in the midst of trying out other ideas while reasoning about a new situation. The case study data in the next section give us the opportunity to consider the factors that can make the blend more or less successful.

## V. CASE STUDY: HOW A KINESIN WALKS

### A. The kinesin task

In In this section we analyze, through the lens of ontological metaphors for energy, one group problem-solving task that asks students to reason about chemical bond energy, and students' work on this task in groups. The kinesin problem was used in both of the two pilot years (2011-13) of the NEXUS/Physics course (with some revisions between the two years) during the weekly discussion section where students work on problems in groups of four. We collected video recordings of two groups during the first year and four groups during the second year; the examples that we analyze here are from the second year.

---
[2] All names are pseudonyms.

Kinesin is a motor protein that "walks" [30] along microtubules to transport cargo within cells. This active transport is powered by the hydrolysis of ATP.[3] In the kinesin task, students are given a "frame-by-frame" description of the kinesin's motion (Figure 2), and are asked to produce energy bar charts to keep track of the energy transformations that take place during this process. They are asked to account for energy conservation in each frame, and are finally asked to discuss what it means to say that a cell "uses ATP to fuel molecular movement" (a statement they might see in a biology class).

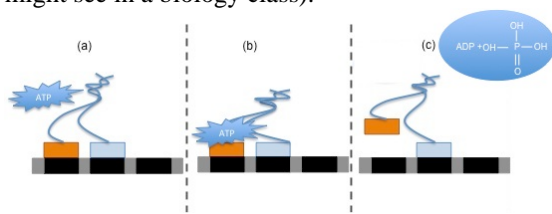

**FIGURE 2.** The picture given to students in the kinesin task, along with a description of the kinesin's motion.

The task was formulated in an open-ended way, and therefore there were many possible approaches the students could have taken (and did take) in creating their energy bar charts. They were explicitly asked to define their system, and were not told which objects to include as part of the system. They were also not told which energies to include in their bar charts, so student groups took different approaches about whether to use chemical energy or potential energy, and whether to consider the chemical/potential energy "of" particular molecules, or of interactions among them.

Though the kinesin task was used only for group discussion and was not graded, we would consider a complete solution to be one that accounted for the kinetic energy of the kinesin, and the changes in chemical (or potential) energy associated with the bonding between the kinesin and the microtubule, between the kinesin and the ATP, and the ATP hydrolysis reaction itself. We would also expect a correct solution to incorporate the correct sign for the changes in energy associated with the formation and breaking of bonds (breaking a bond requires energy to be taken away from some other part of the system). However, the students were not instructed on what level of detail they needed to include. Therefore, it was possible to complete the task in an internally consistent way (at a relatively coarse grain size) by treating all chemical energies as positive (as is done in other settings that use substance-based representations for chemical energy [31]). Nothing internal to the task would necessarily lead the students to reconsider this and shift their representations to using negative energy. This was an unintended consequence of the open-ended task design; while this task was not intended specifically to motivate the need for negative energy, we also expected that students would use negative energy in their bar charts. Some groups did spontaneously use negative energy; others did so only after a suggestion from the TA (and these groups varied in their stances on whether this was something they should have been doing or whether it was a pointless hoop to jump through). Here we examine some of the video data from student groups that were modeling negative energy under these various circumstances.

### B. Phillip's group: Confusion about negative energy

We look first at Phillip's group, working on the energy bar charts portion of the kinesin task. They initially drew all of the bars (including those representing the "chemical energy" associated with the bonds) as positive. The language they use around energy suggests that they are talking about it as a positive substance that can be divided up into smaller pieces. For example, Phillip says "This is like the total energy of the system. It's all chemical right now." Later, when an instructor asks "What's the potential energy here?" Phillip says "100%," and Otis clarifies "Like all of it." (If any of the energies can be negative, it does not make sense to say that "all" of the energy is a particular form, since the kinetic energy could be greater than the total energy as in Figure 3, or the total energy could be zero or negative.)

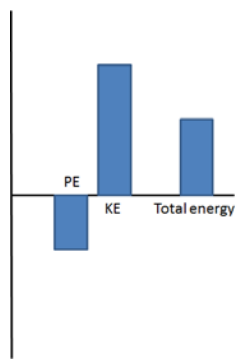

**FIGURE 3.** An example of an energy bar chart in which the kinetic energy is greater than the total energy.

A learning assistant (LA) reminds the group of the potential energy graph that they have seen for chemical bonds, and gets them to say that the energies representing the chemical bonds should be negative. However, they are not entirely convinced that changing their bar graph to include negative bars is necessary. When the TA comes over later and asks them about their decision to make all the bars positive, Phillip responds:

> **Phillip:** *We said absolute value, like the magnitude of the energy.*
> **TA:** *Why did you decide to take the absolute value?*
> **Phillip:** *Because it doesn't really matter later on, because everything else, like this potential, whatever, it just matters where you put the zero.*

Phillip is avoiding negative energy (despite a suggestion to consider it) by making all the energies positive, which is a valid move under some circumstances (possibly including the kinesin task itself). However, he confuses two different methods of making negative quantities positive: translating all the potential energies by a constant amount (moving the zero), and taking the absolute value. While the former method preserves conservation of energy, the latter does not.[3]

In the mathematics education context, Ball [25] writes that "comparing magnitudes becomes complicated. There is a sense in which -5 is more than -1 and equal to 5, even though, conventionally, the 'right' answer is that -5 is less than both -1 and 5. ... Simultaneously understanding that -5 is, in one sense, more than -1 and, in another sense, less than -1 is at the heart of understanding negative numbers."

Similar issues arise in physics, particularly in our interdisciplinary context. In most cases when we talk about negative energy, the "magnitude" is a distraction with no physical significance, since the zero point for potential energy is an arbitrary choice. In those cases, it is obvious that -5 is less than -1 (albeit not always obvious to students). However, in the context of chemical bonds, there is also a sense in which -5 is "more" than -1. A chemical bond with a deeper potential well, associated with a lower (more negative) potential energy, can also be described as a "stronger bond" or "more stable." In chemistry contexts, chemical binding energies are typically reported as positive quantities (absolute values).

Phillip may be activating two different "negative energy can be treated as positive" resources: 1) potential energy is relative, so the zero point can be placed anywhere, 2) "There is a sense in which -5 is more than -1." Each of these resources can be individually useful, but the combination (in the context of energy conservation) leads Phillip and the group to inappropriate reasoning (which will lead to internal inconsistency when they try to keep track of energy conservation) and to resistance to the instructors' interventions. The "potential energy is relative" resource is situated more in the energy-as-location ontology, as we see in Phillip's utterance "where you put the zero." The "-5 is more than -1" resource belongs more to the energy-as-substance ontology: larger negative stuff is more than smaller negative stuff. Thus, this example represents a mixing of substance and location predicates in a way that leads to confusion. This confusion can be manifested both in canonically incorrect results and in internal incoherence.

### C. Peter's group: Productive blending of the substance and location ontologies

Another group working on the same problem starts out talking about energy "stored in the bond," and is unbothered by this idea. As they work through the task and draw their bar charts, they treat all energies as positive, and talk about energy stored in ATP, e.g. "ATP has all the potential energy." But after they overhear the TA saying to another group "...the idea that bound stuff has a negative energy," they quickly reconsider their approach and start incorporating negative energy into their model. When the TA comes over to their group, Peter asks "Would you represent something like the energy that this [kinesin] has while bound to the microtubule as negative energy, 'cause it's like an energy barrier that has to be overcome via the ATP to make it come off?" Tiffany later explains this as "the negative is when energy has to be input to break the bond."

The group classifies which of the energy bars should be positive and negative, and then tries to figure out how to make sure energy is still conserved. They have this discussion, looking at bar charts similar to Figure 4:

---

[3] To illustrate this with a numerical example: Suppose the initial potential energy is -2 and the initial kinetic energy is 5, and the final potential and kinetic energies are -4 and 7. Then the initial and final total energies are both 3, so energy is conserved. Now, if we move the zero of potential energy by 12 so that the initial potential energy is 10, then the final potential energy is 8 (thus all the energies are positive), and the initial and final total energies are both 15, so energy is still conserved. However, if we instead take the absolute value of potential energy, then the initial total energy is 2+5 = 7, and the final total energy is 4+7 = 11, and energy is not conserved.

*Peter:* So what does this have to sum up to?
*Tiffany:* Whatever it starts off at–
*Peter:* Just whatever it started off, ok.
*Tiffany:* Yeah, whatever it starts out at the beginning.
*Zara:* I think it would be negative. The total is [inaudible].
*Peter:* So essentially the well, the net well of the ATP and the bond to microtubule <u>has to equal one big well</u> from the ADP.
*Tiffany:* 'Cause at the end we'll be left with two things. We had the kinetic and the–
*Peter:* But kinetic's up.
*Tiffany:* Yeah.
*Peter:* And the ADP is down. So the ADP has to be *so low* that it's equal to the initial two gaps put together, plus wherever (*Zara:* yeah) the velocity goes. So, ok, so ADP is like waaaaay down. Essentially.
*Zara:* Yeah.
*Peter:* Ok. Got it.

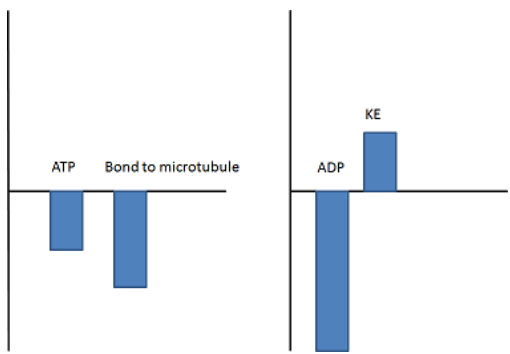

**FIGURE 4.** A reconstruction of the bar graphs drawn by Peter's group.

Peter is doing qualitative arithmetic with the energy bar charts, using positive and negative bars. The bar chart representation is intended to illustrate the conservation of energy by showing that all the bars add up to the same total. But this is only visually obvious when all the bars are positive, so that the total area of all the bars is constant in each frame. In Figure 3, even when the lengths of the bars are adjusted (as the group is negotiating in the transcript clip) so that energy is conserved, the total area of the bars in each frame will not be equal, because some bars represent positive quantities and some represent negative quantities. Therefore, an exclusive substance metaphor (which maps the amount of energy to the amount of bar "stuff") does not work here.

We suggest that Peter is combining the substance and location ontologies for energy, though this is more subtle than in the examples in Part IV. When Peter talks about the two wells adding up to one big well, we code this as an energy-as-substance metaphor (even though the "substance" here represents a negative quantity); he is describing the size of a well as "stuff." But when he says the ADP is "so low" and "waaaaay down," he describes the ADP as being at a vertical location. Finally, the logic that "it's equal to the initial two gaps put together, plus wherever the velocity goes" does not seem to be obviously based in either metaphor; rather, Peter seems to be doing (qualitative) algebra in his head, and mapping it back onto the bar chart representation.

Peter's blended ontology, though it contains the same ingredients, is different from the professor's blended ontology in Part IV. There, the professor consistently used the vertical location metaphor for the energy of the system, and the substance metaphor for changes in the energy of the system. Here, it is more difficult to isolate when each metaphor is used: does ADP **have** a well, or is it **in** a well? It is possible that the use of the metaphors is determined by the type of operation that is being performed: addition of negative numbers is simple enough that it can be visualized with a substance ontology (in the same manner as addition of positive numbers, of which it is just the mirror image), but other operations such as subtraction involving both positive and negative numbers require the location ontology. There is not enough data here to reach a strong conclusion about the exact nature of the blended ontology that Peter uses here. However, he uses this combination of metaphors in this moment to make progress on this energy task. This progress is evident in that he is able to account for the conservation of energy in a way that both matches the canonically correct process and is internally coherent (in contrast to Phillip's group, which struggles to reach this coherence).

Unlike Phillip, who uses a resource associated with one ontology when a resource associated with another ontology would be warranted, Peter uses the two ontologies in complementary ways. In the episodes that we have focused on, Peter's approach is more successful, suggesting that combining the substance and location ontologies is insufficient; the blended ontology needs to have a structure within which the two metaphors can complement each other. Even though both ontologies are in use, they do not collide.

## VI. IMPLICATIONS AND FUTURE DIRECTIONS

Interdisciplinary contexts for teaching physics are becoming more widespread and essential as physics becomes more integrated with the other sciences at both the professional and the educational level. Teaching energy in physics-for-life-sciences contexts, in which chemical reactions are a central phenomenon of study, implies a more primary role for the concept of negative energy in introductory courses. Negative energy furthers the goal of bridging canonical physics models of potential energy (e.g. electrostatic interactions based on Coulomb's Law) with canonical chemistry and biology models of bonding and chemical reactions (e.g. attending to overall energy changes in a reaction [3]). We have argued here that this goal changes the pedagogical considerations and conclusions regarding ontological metaphors for energy that are supported in the instructional context. Instead of focusing on a single ontology for energy, capturing all the relevant characteristics of energy for building this bridge requires a blended ontology.

In the same way that coordinating multiple representations [21] has been shown to be useful in building expertise in both energy and other domains, we suggest that coordinating multiple ontological metaphors can accomplish similar goals in moving towards expertise. We see this productive coordination of ontological metaphors in the data, with examples from experts as well as students who are displaying reasoning that is expert-like to varying degrees. Experts have developed a coherent blended ontology; when students access multiple ontological resources, they have the possibility of coordinating them coherently, or of confusion. When the ontologies are mixed haphazardly, this may lead to confusion, but if the blended ontology has a governing structure, it can be productive. The possibility of confusion has motivated other authors to call for the primary use of the substance ontology in instruction, but in our interdisciplinary context, we call for instructional approaches that help students achieve coherent coordination of ontological metaphors. This may not be necessary or the most effective use of effort in all pedagogical contexts, but in interdisciplinary physics contexts that foreground chemical energy, attention to blended ontologies for energy is worthwhile.

Going forward with this agenda raises a number of practical and theoretical questions, which provide directions for future work. To what extent can student difficulties with the ontology of negative energy be attributed to the specific context of energy, and to what extent do they represent more general difficulties with negative numbers (as documented in the math education literature)? How can the coherent blending of ontological metaphors for energy be explicitly taught? Others [16,19,31,32] have developed representations and activities that can comprise an energy curriculum based on the substance metaphor. Are there representations that can support blending?[33] Or is ontological blending best supported by the coordination of multiple representations, each associated with a single ontological metaphor? The coherent coordination of ontologies requires the development of epistemological resources to determine when it is appropriate to use each metaphor; what pedagogical approaches can support this development?

These issues around ontologies for physical concepts are complicated, and the ontologies that students use are dynamic and arise from multiple sources. Therefore, advising educators to be careful about the metaphors they use in their own speech [34] is neither feasible nor likely to be effective. Conversely, even if experts already use blended ontological metaphors in their speech, we would not expect that their continuing to do so would be sufficient to help students develop blended ontologies, since mere exposure to multiple ontologies is not sufficient to build them into a coherent structure.

The existing work on ontological metaphors for energy has focused on introductory courses, and we have shared that focus, albeit in a specific interdisciplinary course context. However, the "expert" examples that we have presented suggest that blended ontologies for energy may be productive for physicists even in the absence of the interdisciplinary considerations that motivate us. A new direction to explore is the role of blended ontologies for energy in (not necessarily interdisciplinary) physics courses beyond the introductory level.

Implications for researchers include an illustration of the use of the dynamic ontologies framework for making sense of students' reasoning. When this framework is applied to a complex interdisciplinary issue, we see a phenomenon that had not previously been documented within this framework: the productive coordinated use of multiple ontologies in service of a single explanation of a physical phenomenon (as distinct from the ability to access multiple ontologies for the same entity in different situations). This opens up a research agenda to explore ontological blending beyond the energy contexts, both in its general aspects and as applied to other physical phenomena.

## ACKNOWLEDGMENTS

The authors thank Ayush Gupta and the rest of the UMD Physics and Biology Education Research Groups for substantive discussions and feedback. Thanks also to Aaron Eichelberger for some of the video data analysis, to Jessica Clark, Abigail Daane, Ben Van Dusen, and the FFPER Graduate Symposium for helpful comments on an earlier version of this paper, and to all the students who participated in interviews. This work is supported by the NSF Graduate Research Fellowship (DGE 0750616), NSF-TUES DUE 11-22818, and the HHMI NEXUS grant.